\documentclass[article,onecolumn,preprintnumbers,12pt,showpacs,aps,prd,nofootinbib]{article}
\usepackage{a4wide}
\usepackage{fancyhdr}
\usepackage{amsfonts}
\usepackage{mathrsfs}
\usepackage{graphicx}
\usepackage{amssymb}
\usepackage{color}
\usepackage{amsmath}
\usepackage[
       colorlinks=true,
      filecolor=black,
       anchorcolor=blue,
      linkcolor=blue,
      citecolor=red,
      urlcolor=blue,
       linktocpage=true,
        plainpages=false,
        breaklinks=true,
            pdfstartview=Fit
           ]{hyperref}

\usepackage{graphicx}
\usepackage{bm}
\usepackage{float}
\usepackage{color}

\newcommand{\nb}[1]{\color{blue}}

\usepackage[normalem]{ulem}
\usepackage{amsmath}
\usepackage{enumerate}
\usepackage{amsfonts}
\usepackage{yfonts}
\usepackage{subfigure}
\usepackage{psfrag}

\usepackage{epsfig}
\usepackage[latin1]{inputenc}
\usepackage{float}
\usepackage{graphicx}
\usepackage{cancel}
\usepackage{mathrsfs}
\usepackage{amssymb}
\usepackage{amsfonts}
\usepackage{amsmath}
\usepackage{slashed}
\usepackage{placeins}

\def\({\left(}
\def\){\right)}
\def\[{\left[}
\def\]{\right]}
\def\<{\langle}
\def\>{\rangle}
\def\ta{\psi}
\def\tc{\phi}
\def\mc{\lambda}
\def\inter{{}}
\def\[{\left[}
\def\]{\right]}
\def\nn{\nonumber}
\def\dr{\mathrm{d}}

\def\cO{{{O}}}
\def\mM{\mathcal{M}}
\def\nN{n}
\def\nQ{q}
\def\aa{\alpha}
\def\bb{\beta}
\def\bg{R}
\def\Da{\Delta_\Psi^-}

\def\Dpa{\Delta_\phi^-}

\def\bt{\partial\mathcal{M}}
\def\S{{J}}
\def\C{{O}}
\def\Vt{V_{2}}
\def\aa{{\S}_{\Psi}}
\def\bb{{\C}_{\Psi}}

\allowdisplaybreaks

\begin{document}

\title{\bf\large\qquad Engineering Holographic Superconductor Phase Diagrams
}

\author{\small
Jiunn-Wei Chen$^{1,2} $\footnote{E-mail: jwc@phys.ntu.edu.tw},
~~Shou-Huang Dai$^{3} $\footnote{E-mail: shdai.hep@gmail.com},
~~Debaprasad Maity$^{4} $\footnote{E-mail:  debu.imsc@gmail.com},
~~Yun-Long Zhang$^{1} $\footnote{E-mail: zhangyunlong001@gmail.com}\qquad\\
\\
\footnotesize $^1$Department of Physics, Center for Theoretical Sciences, and Leung Center for Cosmology \\
\footnotesize  and Particle Astrophysics, National Taiwan University, Taipei 10617, Taiwan\\
\footnotesize $^2$ Center for Theoretical Physics, Massachusetts Institute of Technology, Cambridge, MA 02139, USA\\  
\footnotesize $^3$Center for General Education, Southern Taiwan University of Science and Technology, Tainan 71005, Taiwan\\
\footnotesize $^4$Department of Physics, Indian Institute of Technology, Guwahati, India\\}

\date{\footnotesize(March 26, 2016)}

\maketitle

\thispagestyle{fancy}
\rhead{\footnotesize{MIT-CTP/4971}}
\renewcommand{\headrulewidth}{0pt}

\begin{abstract}
We study how to engineer holographic models with features of a high temperature superconductor phase diagram. We introduce a field in the bulk which provides a tunable ``doping" parameter in the boundary theory. By designing how this field changes the effective masses of other order parameter fields, desired phase diagrams can be engineered. We give examples of generating phase diagrams with phase boundaries similar to a superconducting dome and an anti-ferromagnetic phase by including two order parameter fields. We also explore whether the pseudo gap phase can be described without adding another order parameter field and discuss the potential scaling symmetry associated with a quantum critical point hidden under the superconducting dome in this phase diagram.
\end{abstract}


\newpage
\lhead{}
\rhead{}

{\small \tableofcontents}

\vspace{1cm}
\section{Introduction}
In the modern condensed matter physics, one of the important areas of research is to understand various phases of quantum matter and their transitions. A prime example of such an experimental system with a plethora of phases is the 
superconducting cuprates. For a conventional (low temperature) superconductor, its many-body ground state and low energy behaviour can be described by the BCS theory \cite{BCS1,BCS2}, which is a weakly coupled theory. However, for high temperature superconductors, their superconductivity appears to originate from strongly correlated many-electron system confined on two dimensional plane due to the special molecular structure of the cuprates.

Over the years many other systems with strong correlation have been discovered, such as the superconducting iron-based compounds known as Pnictides \cite{pnictides}. One of the common features of  these systems is the existence of a superconducting dome in the phase diagram. It has become increasingly evident that, a significant part of the normal phase right above the dome in these systems, e.g. the strange metal phase, can be described by an emergent conformal field theory (CFT). The CFT is scale free at the quantum critical point where some parameter is tuned to a critical value and the temperature is zero \cite{Sachdev2011,Sachdev:2011wg}. Then near the critical point, universal, model independent long range physics exists for models within the same universality class which is governed by the deformed CFT with a few symmetry breaking terms. For example, the strange metal behaviour exists not only in the high temperature superconductors but also in many-spin systems, e.g. heavy fermion metals \cite{Mathur}, near the quantum critical point. The quantum critical point could be hidden or screened by the superconducting dome but revealed by universal low energy behaviours at temperature higher than the dome, such as universal dimensionless transport coefficients which are independent of the microscopic interactions \cite{Sachdev:2011cs}.

Since most of these systems are strongly coupled, conventional perturbative QFT is not applicable. Different models are proposed to explain the mechanism, for example, some systems are believed to accommodate competing symmetry-breaking order parameters in specific range of the physical parameters spanning the phase diagram. 
Based on this idea of competing orders, many new concepts have been introduced, such as the spin fluctuation superglue \cite{moriya}, the resonating valence bond (RVB) gauge approach \cite{anderson,weng} and the SO(5) theory \cite{demler}. In spite of all these new concepts, and their partial success, full understanding of such systems remains elusive. In view of the non-perturbative nature of the problem, AdS/CFT correspondence \cite{Maldacena:1997re,Gubser:1998bc,Witten:1998qj}, a tool developed in string theory, provides a new approach to attack the problem. 

AdS/CFT correspondence, or called the holographic principle, is an duality between a weakly coupled gravity in Anti-di Sitter space (AdS) and a strongly coupled CFT living on the boundary of the AdS space. This duality spawns a spate of research on a certain class of condensed matter-like systems near the quantum critical point. Over the years, the study in the quantum critical property and the intriguing universal low energy predictions indicates that the universality may have some deep connection with the universal transport properties observed in the normal phase (e.g. the strange metal) of the aforementioned class of real systems  \cite{Cubrovic:2009ye,Faulkner:2009wj,Jensen:2010ga}.

AdS/CFT correspondence states that for every particle state in AdS space there exits a dual operator in the corresponding CFT on the boundary. The isometry group of the AdS space is identified with the conformal symmetry of the field theory. This is the symmetry that plays the main role in characterising various quantum phases and their transitions. One of such phases that has recently been extensively studied at finite charge density is called the ``semi-local quantum liquid'' \cite{Iqbal:2010eh}. At low energy this quantum state exhibits different scaling behaviour in time and space. Finite temperature phase transition is realized by considering a charged non-extremal black hole in AdS space. In the extremal limit, the zero temperature quantum phase transition of this system is achieved by varying conformal dimension of the field theory operators. One way to tune the conformal dimension of the strongly coupled operator is by explicitly changing the mass squared of the dual single particle states in a weakly coupled AdS black hole background. However, this way of changing the conformal dimension is not obviously mapped to tuning an experimental parameter from the boundary theory point of view.

Our goal in this paper is to construct a minimum model in bottom-up AdS/CFT that has a similar phase diagram to a high temperature superconducting cuprate. We use a charged black hole background which describes a system of finite temperature and chemical potential. Then we introduce a scalar source $J_{\phi}$ for the boundary field theory as a tuning parameter which arises from the near boundary behaviour of a dual massive scalar field $\phi$ in this background. Then in order to describe the anti-ferromagnetic and superconducting phases, we introduce two massive scalar fields $\{\psi_1,\psi_2\}$. For simplicity, both scalars $\{\psi_1, \psi_2\}$ are neutral with no internal quantum number. By choosing appropriate coupling function between the tuning field $\phi$ and the order parameter fields $\{\psi_1,\psi_2\}$ in the gravity theory, we can design how the effective masses of $\{\psi_1,\psi_2\}$ depend on $\ J_{\phi}$ at $T=0$, then engineer various phase diagrams in the $\{ J_{\phi}, T \}$ plane, including those similar to real high temperature superconducting cuprates. One can easily generalise this setup to either $U(1)$ charged field with superconducting condensation or different kind of anti-ferromagnetic models,
without changing the qualitative behaviour of the phase diagram.
Study along the similar line has been reported very recently in \cite{Kiritsis:2015hoa},
where a different system with an extra $U(1)$ gauge field on the gravity background is introduced to represent a chemical potential, as an external tuning parameter for the phase transition in high temperature superconductors. This is an example of modeling the phase transitions in reality due to varying a tunable doping parameter. In the global AdS background, the holographic quantum phase transitions and interacting bulk scalars are also studied in \cite{Chaturvedi:2014dga}.

This paper is organized as follows. In Section \ref{SecSetup}, we introduce the setup and review  the holographic properties of semi-local quantum liquid. In Section \ref{SecPhase}, we discuss the engineering of phase diagrams in holography to obtain phase diagrams that has the features of high temperature superconducting cuprates. 
In Section \ref{scaling}, we discuss the scaling symmetries near quantum region that is not manifest in our models.
Our results are summarized and discussed in Section \ref{SecCon}.

\section{Background gravity}
\label{SecSetup}

In this section we review some basic properties  of the gravitational background. It has a charged black hole in asymptotic AdS background to describe a system of finite temperature and chemical potential. 
 
Following the work of \cite{Iqbal:2011aj,Iqbal:2011ae}, we introduce the action of the Einstein gravity with Maxwell and other matter fields (which will represent some order parameters in the dual system) in the $3+1$ dimensional bulk:
\begin{align}\label{action}
S_{bulk}&=\int\dr^4 x \sqrt{-g}\[\frac{1}{2\kappa^2}\(R-2\Lambda\)-\frac{L^2}{2\kappa^2}\frac{1}{g^2_F}F^2\]+S_{b.t.}\nn\\
&+\int\dr^4 x\sqrt{-g}\(\mathcal{L}_{M}\)+S_{c.t.}, 
\end{align}
where $S_{b.t}$ and $S_{c.t}$ denote the boundary terms and the counter terms respectively. The coupling $\kappa^2= {8\pi G_N}/{c^4}$ is of dimension $[L]^2$, related to the Newton's constant $G_{N}$ and the speed of light $c$. The negative cosmological constant is given by $\Lambda=-{3}/{L^2}$, where $L$ is the AdS radius. The effective gauge coupling $g^2_F$ of the Maxwell term is dimensionless.

If we exclude the matter field for now, the Einstein's and Maxwell's equations of motion are 
\begin{align}
R_{\mu\nu}-\frac{1}{2}R g_{\mu\nu}+\Lambda g_{\mu\nu}
 &=\, \frac{2L^2}{g^2_F}\[F_{\mu\lambda}F^{\lambda}_{~\nu}-\frac{1}{4}F^2g_{\mu\nu}\],\nonumber\\ \nabla_\mu {F^{\mu}}_{\nu} &=\, 0.
\end{align}
The vacuum solution we consider is a charged black brane,
\begin{align}\label{solution}
d s^2& = \frac{r^2}{L^2}\[-f(r)d t^2+d x^2+d y^2\]+\frac{L^2}{r^2}\frac{d r^2}{f(r)},\nn\\
F &= g_F \frac{\nQ}{L^2}\frac{r^2_h}{r^2}\dr r\wedge \dr t,
\end{align}
where $\nQ$ is the charge density of the black brane.
The factor $f(r)$ and electric potential $A_t(r)$ are
\begin{align}\label{fandA}
f(r)& =1-\(1+{\nQ^2}\)\frac{r_h^3}{r^3}+{\nQ^2}\frac{r_h^4}{r^4},\qquad \nonumber \\
A_t(r)&=g_F\frac{\nQ\,  r_h }{L^2}\(1-\frac{r_h}{r}\).
\end{align}
We choose the gauge $A_r=0$,
and $r_h$ is the outer horizon of the charged black brane, satisfying $f(r_h)=0$.
According to the standard AdS/CFT dictionary, the temperature and the entropy density of the boundary semi-local quantum liquid are identified as those of the black brane,
\begin{align}\label{temperature}
T=\frac{3}{4\pi}\frac{r_h}{L^2}\[1-\frac{{\nQ^2}}{3}\],\qquad
s=\frac{2\pi}{\kappa^2}\frac{r_h^2}{L^2}.
\end{align}
The chemical potential and the charge density are given by \cite{Iqbal:2011aj},
\begin{align}\label{chemical}
 \mu_{\nQ} = g_F \nQ \frac{r_h}{L^2},~~\qquad \nN_{\nQ}=\frac{2\nQ}{\kappa^2g_F}\frac{r_h^2}{L^2}.
\end{align}
The temperature in \eqref{temperature} can also be expressed in terms of the chemical potential and the position of outer horizon,
\begin{align}\label{eq8}
T(\mu_\nQ,r_h)=\frac{3}{4\pi} \frac{r_h}{L^2} \[1-\frac{1}{3}\frac{L^4  }{g_F^2 }\frac{ \mu_\nQ^2 }{ r_h^2}\].
\end{align}
According to the above expression, there are two ways to achieve zero temperature. One is to reduce to pure AdS space, i.e. $r_h=0$ and $\mu_\nQ=0$ such that $T(0,0)=0$, which also implies vanishing charge density. But we are interested in systems with finite density, and therefore we achieve zero temperature by taking the extremal limit of the charged brane $\mu^*_\nQ=\sqrt{3}g_F  {r_h}/{L^2}$ with nonvanishing $r_h$, such that $T\(\mu^*_\nQ, r_h\)=0$.

The basic guiding principle of AdS/CFT is based on the symmetries on both sides of the correspondence. Scaling symmetry is
one of those larger symmetries which plays an important role in understanding the low energy behaviour
of the system under consideration in terms of relevant operators of definite scaling dimension.
One such interesting system dual to the charged black hole in Anti-de sitter(AdS) space is known as ``semi-local quantum liquid'' \cite{Iqbal:2010eh}. Following reference \cite{Hartnoll:2009sz}, we review the scaling symmetries for the semi-local quantum liquids. Our solution is parameterized in terms of two independent parameters $\nQ$ and $r_h$, while the appropriate physical parameters we consider are temperature $T$ and the chemical potential $\mu_\nQ$. Therefore, in terms of those
thermodynamic variables, the equation of state of the dual field theory of the charged black brane turns out to be
\begin{align}
\nN_{\nQ}\(\mu_\nQ,T\)=\frac{4 \pi}{3\kappa^2}\frac{ L^2}{g^2_F} \mu_\nQ T\(1+\sqrt{1+\frac{3}{4\pi^2 g_F^2}\frac{\mu_\nQ^2}{T^2}}\). \label{EOS}
\end{align}

The equations of motion respect two types of scaling symmetries \cite{Gubser:2008px}. The first type is the global scaling 
\begin{align}
L\rightarrow a L,
\quad \{r,t,x,y\}\rightarrow a\{r,t,x,y\},\quad \kappa \rightarrow a^2\kappa^2,
\end{align}
which rescales the metric $ds^2 \rightarrow a^2 ds^2$, and the physical quantities are scaled accordingly,
\begin{align}
\{T,\mu_\nQ\}&\rightarrow a^{-1}\{T,\mu_\nQ\},\qquad 
\{s, n_\nQ\}\rightarrow a^{-2}\{ s, n_\nQ\}.  \label{scaling1}
\end{align}
One can make use of it to scale away $L$ to unit in the physical quantities,
and we will take $L=1$ in our numerical analysis, meaning every length is measured in units of the AdS radius. The second type is
\begin{align}\label{scalingx}
 \quad r \rightarrow \lambda r,
 \quad  \{ t,x,y \} \rightarrow  \lambda^{-1} \{t,x,y\},
\end{align}
which leaves $ds^2$ invariant
and the physical quantities, 
\begin{align}
\{r_h,T,\mu_\nQ\}&\rightarrow \lambda\{r_h,T,\mu_\nQ\},\qquad 
\{s, n_\nQ\}\rightarrow \lambda^2\{ s, n_\nQ\}.  \label{scaling1}
\end{align}
Via this type of scaling, one can initially set horizon size to unit, i.e. $r_h=1$, for the convenience of computation. 
In the following numerical analysis, we will set in the beginning $r_h=1$ for convenience, 
and retain $r_h$ in the equations and diagrams through rescaling it back to the required size afterwards.
On the other hand, one can redefine the scaling invariant charge density
$\tilde\nN_{\nQ} = (\nN_{\nQ}g_F\kappa^2) /{ (T^2L^2)} $ and chemical potential $ \tilde\mu_\nQ= {\mu_\nQ}/(T g_F)$,
such that the rescaled equation of state from Equation \eqref{EOS}  is independent of the temperature $T$,
\begin{align} \label{nqt}
\tilde\nN_{\nQ}\(\tilde\mu_\nQ\)=\frac{4\pi }{3} {\tilde\mu_\nQ} \(1+\sqrt{1+\frac{3}{4\pi^2}  {\tilde\mu_\nQ^2}}\).
\end{align}
Similar scale invariant equation of state in the quantum critical region has been observed in experiments \cite{ChinCheng1,ChinCheng}. 
\section{Phase diagram engineering}
\label{SecPhase}

In this section, we would like to construct a minimum holographic model in the bottom-up approach that has a similar phase diagram to a high temperature superconducting cuprate (see, e.g., Figure 1 of \cite{HTCSC1}). As we have already discussed above, at low temperature, distinct phases can be obtained by tuning the doping parameter.
In order to understand the basic mechanism of quantum phase transition near zero temperature, we consider interacting
order parameter fields. 

According to the standard AdS/CFT dictionary, the chemical potential $\mu_q$, conjugate to the charge density $n_q$, is dual to a bulk gauge field $A_{\mu}$, such that $\mu_q$ and $n_q$ are encoded in the non-normalizable and normalizable modes of the asymptotic behavior of $A_{\mu}$ respectively. Similarly, in our model in this section, the doping parameter should be dual to a bulk scalar field $\phi$. In the asymptotic solution of this so-called ``tuning field'' $\phi$, the non-normalizable mode is dual to the source $J_{\phi}$ on the boundary, interpreted as the doping parameter since it is an intensive quantity, just like the role of the chemical potential as the non-normalizable mode of the asymptotic $A_{\mu}$. The normalizable mode of $\phi$, on the other hand, is dual to the expectation value of the conjugate variable to the doping parameter, which we do not specify.

We consider two order parameter
fields to be neutral scalar fields $\psi_1, \psi_2$ in AdS bulk. We also conjecture that the controlling parameter of our system is dual to another
neutral field $\phi$ which is coupled to $\psi_1, \psi_2$ with a certain degree of fine tuning, such that we can reproduce the experimental phase diagram.

Our goal is to understand the phase diagram and the scaling behavior near the quantum critical point in such a system. 
In order for the two order parameters to be controlled by tuning the external parameter, it requires $\psi_1$, $\psi_2$ interact with the tuning field $\phi$ in some non-linear way.  Therefore, we introduce the following minimal Lagrangian density
\begin{align}\label{LagrangeD}
 \mathcal{L}_{M}&=  \sum_{i=1,2} \mathcal{L}_{\ta_i} + \mathcal{L}_{\tc}+\mathcal{L}_{int} ,
\end{align}
where
\begin{align}
g_M^2\mathcal{L}_{\ta_i}&=-\frac{1}{2}\(\partial\ta_i\)^2-V(\ta_i),\quad V(\ta_i) =\frac{1}{2}m_i^2\ta_i^2+\frac{1}{4}\mc_i \ta_i^4, \\
g_M^2\mathcal{L}_{\phi}&= -\frac{1}{2}\(\partial\tc\)^2-V(\tc),\qquad
V(\tc)=\frac{1}{2}m_\tc^2\tc^2+\frac{1}{4}\mc_\tc \tc^4,
\end{align}
with $g^2_M$ indicate the coupling constant,  $m_i^2, m_\tc^2<0$ and $\lambda_i, \lambda_\tc>0$.
The interaction terms between $\psi_1$, $\psi_2$ and $\phi$ are given by
\begin{align}
g_M^2\mathcal{L}_{int}=&-\frac{1}{2}\inter\sum_{i=1,2}{F_i(\tc)}\ta_i^2, \label{crossint}
\end{align}
where the detailed form of the coupling function $F_i(\tc)$ will be given later. Different $F_i(\tc)$ implies different ways that the condensation of $\psi_1$ and $\psi_2$ are controlled by $\phi$ via shifting their effective masses. Consequently, different phase structures arise.
We will work in the probe limit of the scalar fields, namely ${2\kappa^2}/{g_M^2}\rightarrow0$.

The equations of motion for the scalar fields turn out to be
\begin{align}\label{ScalarEom}
0&=r^{-2}  \partial_r\[r^4 f(r) \partial_r{\ta_i}\]-\[m_i^2 +  \inter F_i(\tc)\] \ta_i- \mc_i\ta_i^3,\\
0&= r^{-2}   \partial_r\[r^4 f(r) \partial_r{\tc}\]-m_\tc^2\,\tc
-\frac{1}{2}\inter\sum_{i=1,2}{F'_i(\tc)}\ta_i^2 -\mc_\tc\tc^3.
\end{align}
It is clear from these equations that, at zero temperature limit, the tuning field $\tc$ shifts the effective mass of $\psi_i$ near the horizon from $m_i^2$ to $\tilde m^{2}_i=m_i^2 + \inter F_i\(\tc(r_h)\)$. This is similar to the case in \cite{Iqbal:2010eh}.

At finite temperature, the near horizon expansions of the scalar fields are
\begin{align}
\ta_i(r)&= \ta_i(r_h)+\ta'_i(r_h)(r-r_h)+...,\\
\tc(r)&= \tc(r_h)+\tc'(r_h)(r-r_h)+...,
\end{align}
and solving the equation of motion leads to
\begin{align}
\ta'_i(r_h) &= \[m_i^2 \ta_i(r_h)+ \mc_i \ta_i(r_h)^3\]/{(4\pi T)},\\
\tc'(r_h)&= [m_i^2 \tc(r_h)+ \mc_i \tc(r_h)^3  +\frac{1}{2}\inter\sum_{i}{F'_i(\tc)}\ta_i^2]/{(4\pi T)}.
\end{align}
These formulae are used to evaluate the scalar field in the numerical analysis once the boundary values on the horizon are specified.
 On the other hand, the asymptotic forms are given by
\begin{align}
\ta_i(r)&\rightarrow \frac{J_i}{r^{\Delta^-_i}}+\frac{\cO_i}{r^{\Delta^+_i}},\qquad  \Delta_{i}^{\pm}=\frac{3}{2}\pm\sqrt{\frac{9}{4}+m^2_i  }~ ,
\label{psii}\quad\\
\tc(r)&\rightarrow \frac{\S_\tc}{r^{\Delta^-_\tc}}+\frac{\C_\tc}{r^{\Delta^+_\tc}},\qquad  \Delta_{\tc}^{\pm}=\frac{3}{2}\pm\sqrt{\frac{9}{4}+m^2_\tc }~ . \label{phiasym}
\end{align}
Once can also read out the scaling dimension of the sources $J_i, J_\phi$ and operators  $O_i, O_\phi$ according to \eqref{scalingx}, 
that
\begin{align}\label{scalingJO}
\{J_i, J_\phi\} \rightarrow \{\lambda^{\Delta_i^-} J_i, \ \lambda^{\Delta_\phi^-} J_\phi\},\qquad
\{O_i, O_\phi\}\rightarrow \{ \lambda^{\Delta_i^+} O_i, \ \lambda^{\Delta_\phi^+}  O_\phi\}.
\end{align}

In our analysis, we choose the standard quantization on the boundary for the asymptotic ${\ta_i}$, that set the boundary source $J_i=0$.
Therefore we do not need to consider the correction term in asymptotic $\psi_i$ due to $\psi_i^4$ self-interaction in \eqref{phic}. 
The cross interaction $F_i (\phi)\psi_i^2$ will not incur corrections to either asymptotic $\psi_i$ or $\phi$ due to the same reason. The $\phi^4$ interaction  induces correction to the asymptotic $\phi$ solution, but we will constrain the mass of $\phi$ to $m_{\phi}^2 <-27/16$ in our analysis, so that the correction contributes at the sub-leading order in the normalizable mode and can be neglected at asymptotic infinity.

In the following, we present two different models, with different sets of $\{m_i^2$, $m_{\phi}^2$, $\lambda_i$, $\lambda_{\phi}\}$ parameters and the coupling functions $F_i (\phi)$.
As expected, tuning the source parameter $J_{\phi}$ (which is the non-normalizable mode of $\phi$), leads to different phase diagram.
Such a phenomena have been observed in the phase diagrams of high temperature superconductors.

To check the thermodynamic stability of any particular phase, we need to examine the free energy of the
condensed state and normal state in the ordered phase (see more details in Appendix \ref{probescalar}).
The free energy densities with and without $\psi_i$ in these  states are given by
\begin{align}
 \frac{\Delta\Omega}{V_{(2)}}  &=  \frac{\Delta\Omega_\phi}{\Vt}-  \frac{1}{4 g_M^2}\int_{r_h}^{\infty} \dr{r}\frac{r^2}{L^2}
 \sum_i \[ \mc_i\ta_i^4+ \phi \inter F_i'(\tc) \ta_i^2  \], \label{freeenergyd1}\\
\frac{\Delta\Omega_\phi}{\Vt} &= -\frac{\lambda_{\phi}}{4 g_M^2 }\int_{r_h}^{\infty}\dr r \frac{r^2}{L^2}\({\phi}^4\),\label{freeenergyd}
\end{align}
where the free energy coming from the pure black brane background has been subtracted. In the subsequent subsections, we will study two possible phase diagrams by tuning the temperature and the source parameter $J_{\phi}$ and checking the free energies to shed light on the possible ground states for a given value of $\{J_{\phi}, T\}$.

\subsection{Model I with positive dopping parameter}
\label{phi4}
In this model, we engineer a phase diagram with two condensed phases mimicking the anti-ferromagnetic phase and the pseudo gap phase. In the Lagrangian density \eqref{LagrangeD}, we choose the following coupling functions $F_1 (\phi)$ and $F_2 (\phi)$, 
\begin{equation}
  - F_2(\phi) = \phi^2 - \frac{5 }{24} \phi^4 \cong F_1(\phi)  . \label{tuningF}
\end{equation}
To avoid the instability of the scalar fields due to the unboundedness from below in the potential in (\ref{crossint}) and (\ref{tuningF}), we can add a small $\phi^6$ term to $F_1(\phi)$, e.g. $F_1(\phi) = \phi^2- 5 \phi^4/24+ \phi^6/100$.
This will only change the full phase diagram slightly. Our choice is tuned to that particular form, so that we can construct a phase diagram which mimics the high temperature superconductivity.
Moreover, this choice also reduces the complication in numerical analysis. We also choose the following set of parameters:
\begin{align}
m_1^2 &= -2.1, \quad \lambda_1=2,\nn \\
m_2^2&= -1.0,\quad \lambda_2=2,\\
m_\phi^2&= -1.8,\quad \lambda_\phi=1/3. \nn
\end{align}
Together with the boundary condition for $\psi_i$ at asymptotic infinity such that $J_i = 0$, the correction terms due to the self- and cross- interaction terms drop out, as explained below equation (\ref{scalingJO}).

\begin{figure}[H]
\begin{center}
\includegraphics[width=0.6\textwidth]{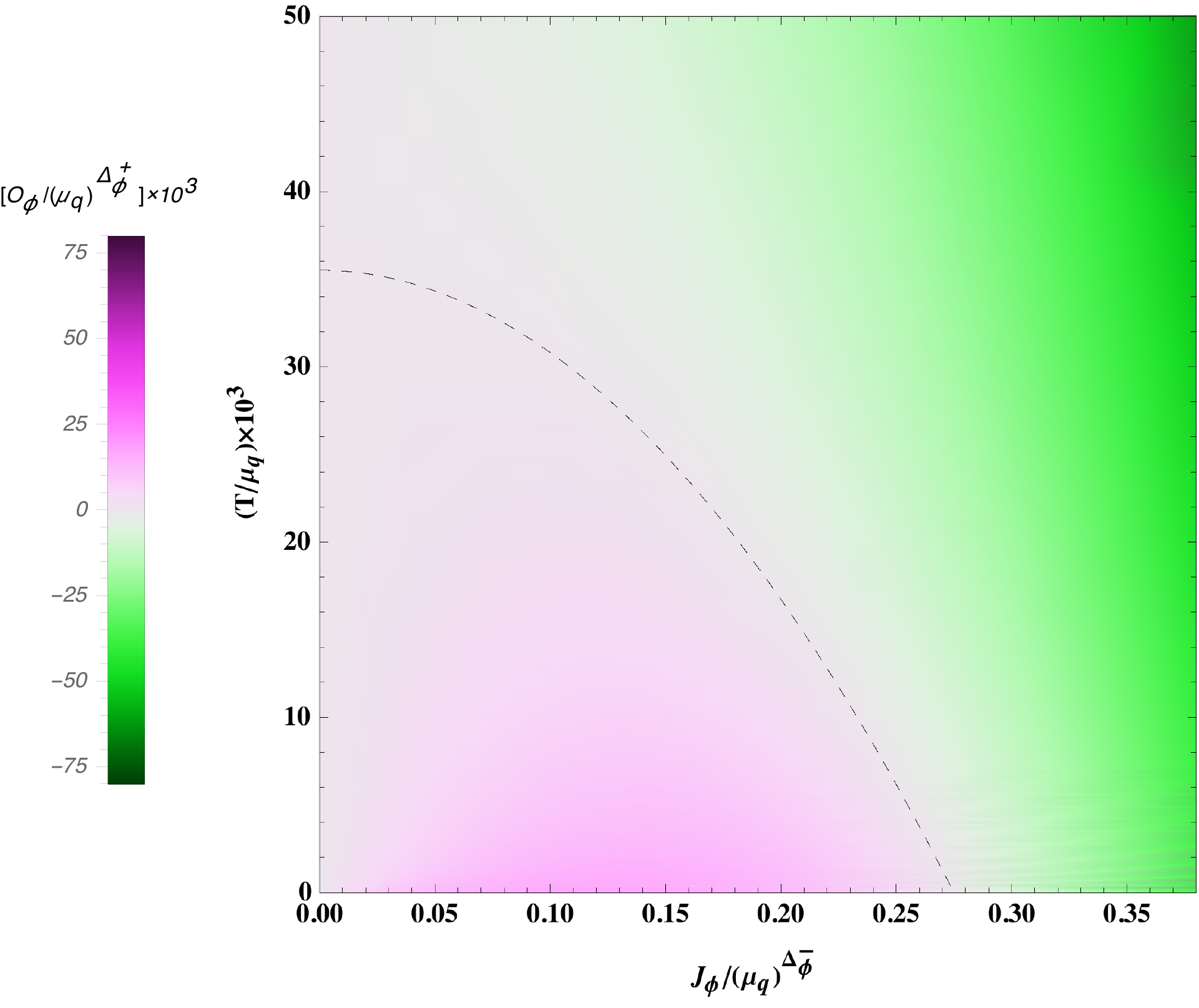}
\caption{The density plot for expectation value of $\cO_{\phi}$ versus the rescaled temperature $T/\mu_q$ and the rescaled doping parameter (the source) $J_{\phi}/(\mu_q)^{\Delta_\phi^-}$, both by appropriate power in the chemical potential $\mu_q$. The dotted line is where $ \cO_{\phi}$ vanishes. Note that $ \cO_{\phi}$ is non-vanishing in general. And the relation between $ \cO_{\phi}$ and $J_{\phi}$ is similar to the ``equation of state'' of the doping matter dual to scaler field $\phi$. }\label{Fullcs2}
\end{center}
\end{figure}

In Figure \ref{Fullcs2}, we firstly plot the expectation value of the dual operator $\cO_{\phi}$ with colour gradient on the rescaled $\{J_{\phi}, T \}$ plane. We only consider the $J_{\phi}$ parameter over a small positive range in this diagram. Physically, $ {\cO}_{\phi} $ is a thermodynamical variable conjugate to the source parameter $J_{\phi}$, their relation is similar to the ``equation of state'' of the doping field $\phi$. 
The location of vanishing $ \cO_{\phi} $ varies with $J_{\phi}$, indicated by the dashed line in the figure. In terms of the order parameter $ {\cO}_{\phi} $,  there is no phase transition across the dashed line. 
It would be very interesting if there were a phase transition in $ {\cO}_{\phi} $ as it would provide a way to describe the pseudo gap as we will discuss in Sec.~\ref{scaling}.

\begin{figure}[H]
\begin{center}
\includegraphics[width=0.75\textwidth]{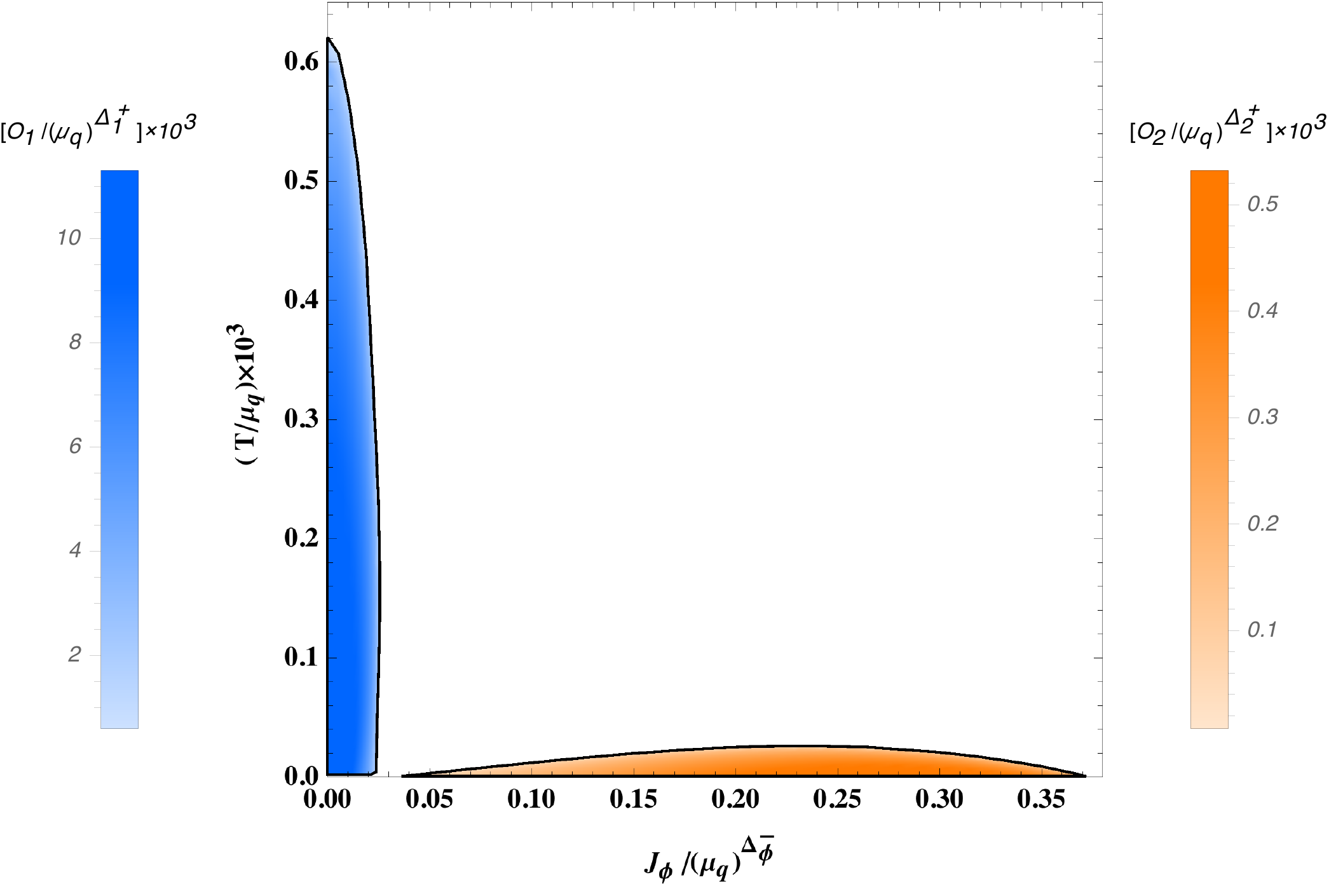}
\caption{
The phase diagram of model I with positive dopping parameter.
It is the density plot for phase 1(blue) and phase 2 (orange) with coupling functions $-F_2(\phi)=\phi^2- 5\phi^4/24 \cong F_1(\phi) $.
}\label{Fullphase2}
\end{center}
\end{figure}

Now we turn on the fields $\psi_1$ and $\psi_2$,
the phase diagram of this system is presented in Figure \ref{Fullphase2}. It contains two distinct ordered phases, characterized by non-trivial values of $ \cO_1 $ and $ \cO_2 $ respectively, besides the normal one.
This diagram is obtained by setting the sources of $\psi_1$ and $\psi_2$ on the boundary to be zero, but turning on the boundary source $J_{\phi}$ of scale field $\phi$ and the background temperature.

In this phase diagram, we only consider a small range of positive rescaled $J_{\phi}$. The ordered phase 1 (in blue color) occurs at low $J_{\phi}$, and extend up to higher temperature, while the dome-shape ordered phase 2 (in orange color) covers the low temperature region down to $T=0$, over a finite range of positive $J_{\phi}$. This feature is qualitatively similar to that of the cuperate superconductors.
If one wish to create a more realistic model, and identify the ordered phase 1 (characterized by non-trivial $ {\cO}_{1} $ )  near $J_{\phi} =0$ to the anti-ferromagnetic phase of the cuperates, one needs to generalize $\psi_1$ to be one component of the $SU(2)$ multiplet \cite{Iqbal:2010eh}. This in principle requires to include other components of $SU(2)$ into the Lagrangian, but only allows $\psi_1$ to condense. Moreover, the order parameter fields and also be promoted to be charged under $U(1)$. Since the goal of this paper is to explore the possibility of forming different phases via the tuning field $\phi$ and the coupling function $F_i (\phi)$, the study of the more realistic model can be left for future work.

\begin{figure}[H]
\begin{center}
\includegraphics[width=0.55\textwidth]{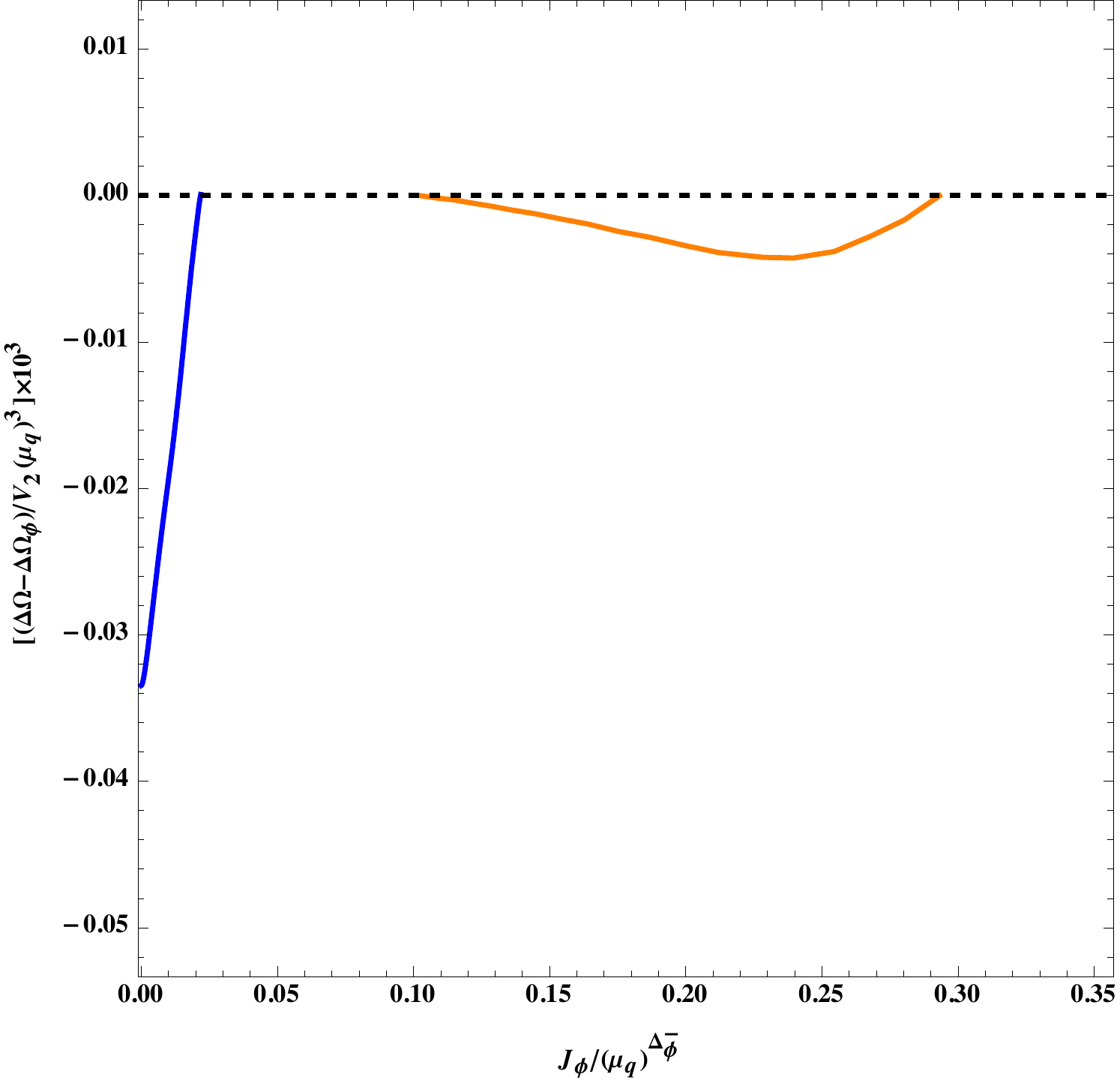}
\caption{the free energy density difference between the ordered and normal phase. The rescaled temperature is choosen to $(T/\mu_q) \simeq 0.014 \times 10^{-3}$. The dashed black line stands for the baseline of the free energy of the normal phase with only $\phi$ condensate. The blue and orange lines correspond to the free energy difference of phase 1 and phase 2, respectively.}\label{free2}
\end{center}
\end{figure}

Figure \ref{free2} shows that the ordered phase 1 and phase 2 are indeed thermodynamically preferred, by comparing the free energy density of the solutions with and without the condensate in the range of $J_{\phi}$ in those phases at a fixed temperature.
The difference of free energy density is based on formulas  \eqref{freeenergyd1} and \eqref{freeenergyd},
and here we only show the free energy density difference $\Delta\Omega-\Delta\Omega_\phi$ at a particular value  $(T/\mu_q) \simeq 0.014 \times 10^{-3}$. The blue and orange lines correspond to the free energy difference of phase 1 and phase 2, respectively.
And the dashed black line stands for the baseline of the free energy of the normal phase with only $\phi$ condensate.

\subsection{Model II with competing orders}
\label{phi2}
In this model, we have a region with two condensates coexisting which provides a model to study physics of competing orders. The phase diagram is also similar to the region with a superconducting dome and a pseudo gap phase. We start with the following simple
choices of parameters in the Lagrangian density \eqref{LagrangeD},
\begin{align}
m_1^2 &= -1.5, \quad \lambda_1=2,\quad F_1(\phi)=\phi(\phi+2) \\
m_2^2&= -1.9,\quad \lambda_2=2, \quad  F_2(\phi)=\phi^2/2\\
m_\phi^2&= -1.5,\quad \lambda_\phi=0.
\end{align}

\begin{figure}[H]
\begin{center}
\includegraphics[width=0.75\textwidth]{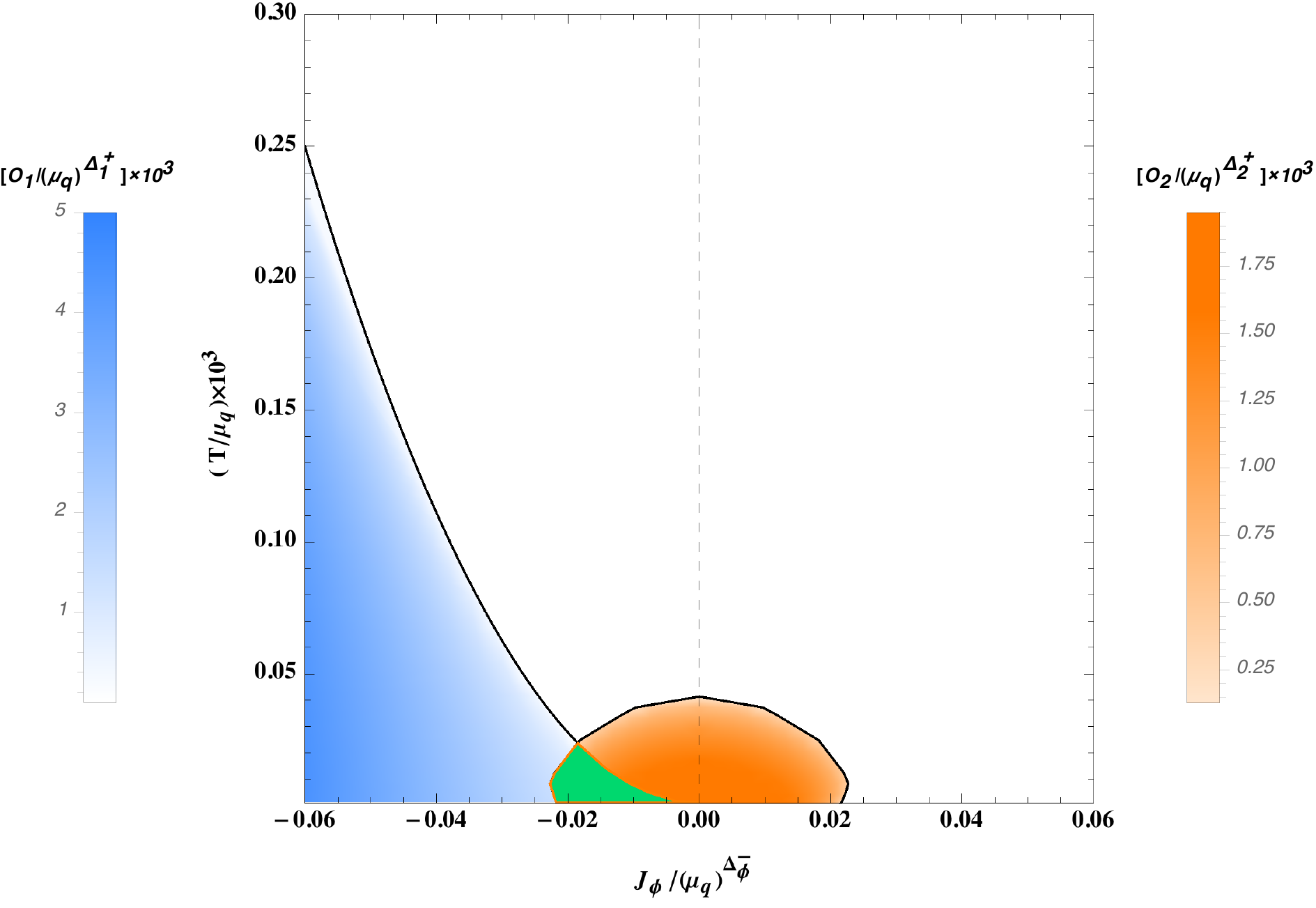}
\caption{
The phase diagram of model II around a natural quantum critical point.
It is the density plot phase 1(blue) and phase 2 (orange) with coupling functions $F_1(\phi)=\phi(\phi+2),~
F_2(\phi)=\phi^2/2$.
The green parts are the overlap region.
}\label{Fullphase1}
\end{center}
\end{figure}

We define the normal phase of the our system with all the order parameter fields $\psi_i=0$, which
is to say that there is no condensation of the corresponding dual field theory operator. Therefore,
the system is in completely symmetric phase.
We numerically study how different possible phases and their transitions are occurring as we go towards the zero temperature
limit for different values of the source parameter $J_{\phi}$ at the boundary field theory.
In this case, we assume the tuning parameter taking both negative and positive values,
which mimics the electron and hole doping into the system for high temperature superconductivity.
Therefore, the $J_{\phi} =0 $ can be identified as the quantum critical point where the transition temperature
is zero. 
The other motivation of the choices of parameters in this model refer to the phase diagrams in \cite{Gauntlett:2009bh},
which is related with at zero source $J$ point through scaling symmetries.
But for our specific choices of parameters, we see from the phase diagram \ref{Fullphase1}, that
the quantum critical point is covered by a dome with non-zero condensation of ${\cO}_{2} \ne 0$,
which normally happens in real physical systems. One also notices that there exists an overlapping phase
which we left for future studies.


\begin{figure}[H]
\begin{center}
\includegraphics[width=0.55\textwidth]{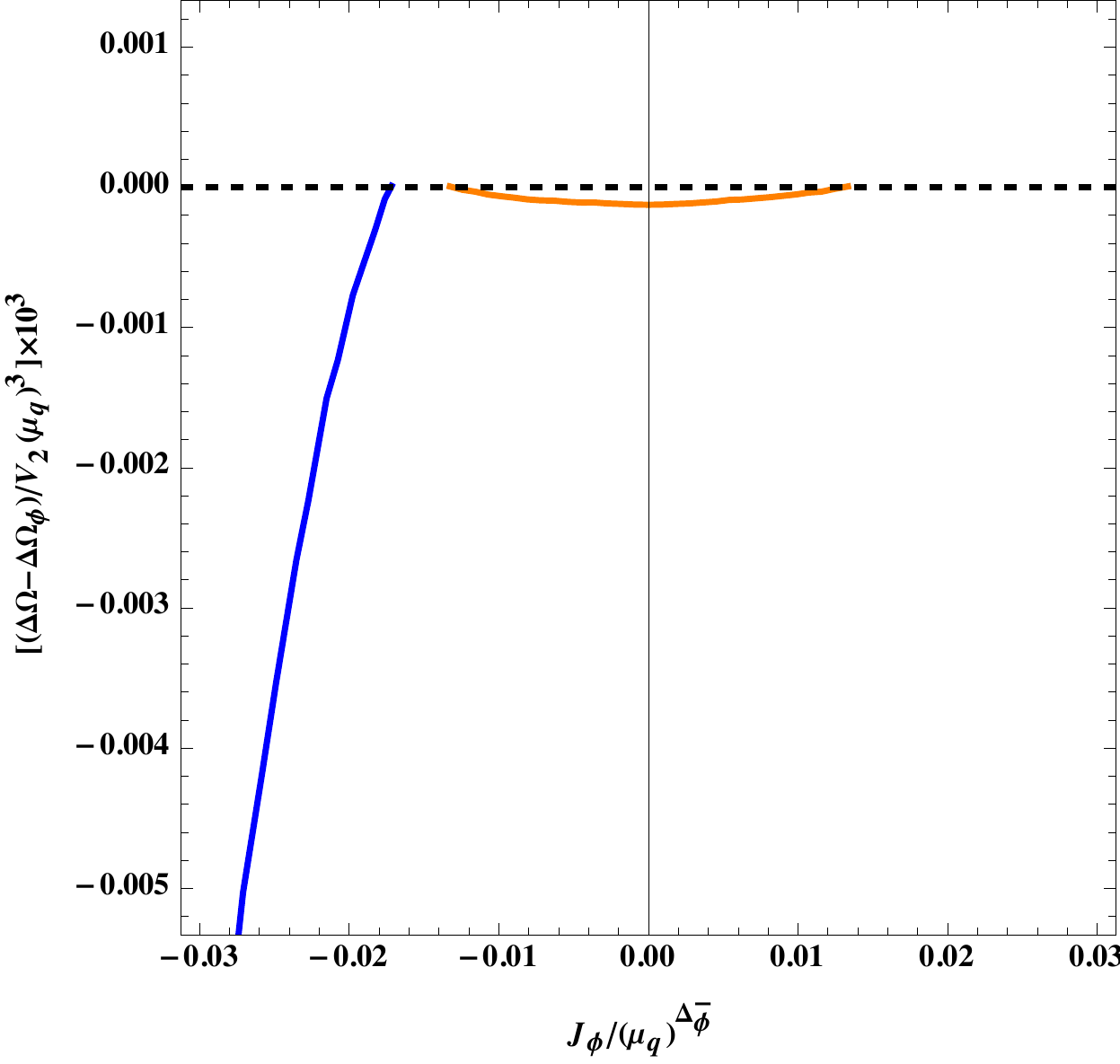}
\caption{The energy difference between the ordered and normal phase 
in Figure \ref{Fullphase1} at  $(T/\mu_q)\times 10^3 \simeq 0.028$.
The dashed black line stands for the baseline of the free energy of the normal phase with only $\phi$ condensate. The blue and orange lines correspond to the free energy difference of phase 1 and phase 2, respectively
}\label{free1}
\end{center}
\end{figure}

Once we numerically compute various condensations for the different phases, one needs to compare the free energy among various phases
in Figure \ref{Fullphase1}.
As we have calculated the free energy for various phase, in the Figure \ref{free1} we have  plotted
them using the expressions \eqref{freeenergyd1} and \eqref{freeenergyd}.
We show the free energy density difference $\Delta\Omega-\Delta\Omega_\phi$ at a particular value  $(T/\mu_q) \simeq 0.028\times 10^{-3}$.
The blue and orange lines correspond to the free energy difference of phase 1 and phase 2, respectively.
And the dashed black line stands for the baseline of the free energy of the normal phase with only $\phi$ condensate.

One notices that in the region $J_{\phi} < 0$, $\Delta \Omega -\Delta \Omega_{\phi}<0 $. Therefore, the
ground state of the system will be in the phase with ${\cO}_{1} \ne 0$. On the other hand near $J_{\phi} =0$ region,
we found that the free energy $\Delta \Omega$ becomes almost comparable to $\Delta \Omega_{\phi}$, as one sees from the
Figure \ref{free1}. But still ${\cO}_{2} \ne 0$ is preferable near the $J_{\phi}=0, T =0$.
The green region is the overlap region, and usually there exist some competing and coexistence orders.
see references e.g. \cite{Basu:2010fa}-\cite{Nie:2013sda}.
This region is not the main purpose of our model, and it would be interesting to explore more on this issue in future work.

\section{Towards a more realistic phase diagram}
\label{scaling}

To make the holographic high temperature superconductor phase diagram more realistic, we need at least the following improvements: (a) realistic condensates for the anti-ferromagnetic and superconducting phases, (b) a pseudo gap phase, (c) scaling symmetry associated with the screened quantum critical point below the superconducting dome. 

For (a), it is not difficult to replace $\psi_1$ of model I by a condensate that has $SU(2)$ components to make it a realistic anti-ferromagnetic phase \cite{Iqbal:2010eh,Cai:2015mja}.
 It is also not difficult to replace $\psi_2$ of the same model by a complex scalar field couples to the bulk $U(1)$ gauge field in Equation(\ref{action}) to have a S-wave superconductor, or different kinds of holographic P-wave superconductors \cite{Hartnoll:2008vx}-\cite{Cai:2015cya}.
 However, experimentally, the condensate for the superconductor is of D-wave. Naively, one might expect that this can be achieved by just employing a symmetric traceless second rank tensor as the order parameter field in the bulk. However, this naive construction has more components than needed which need to be removed in a general covariant way. The fact that this field is charged and massive in the AdS background makes things even more complicated  \cite{Chen:2010mk}. This is the difficulty of constructing high spin field theory. It is a long standing problem and is not yet resolved despite lots of efforts \cite{Benini:2010pr,Hartnett:2012np,Kim:2013oba}.    

In the following subsections we discuss the missing pseudo gap phase and scaling symmetry in our phase diagram Figure \ref{Fullphase2} or Figure \ref{Fullphase1}, which is based on  the models presented in Section \ref{SecPhase}.

\subsection{Pseudo gap phase}

The definition of a pseudo gap phase is where fermion spectral function has a gap, but the order parameter is zero. For cuprate superconductors, the pseudo gap phase occurs at the temperature  $T*$  above the superconductor phase transition $T_c$ ($T* > T_c$), where the superconducting order parameter vanishes, but the gap in the fermionic spectral function remains finite. 

A holographic pseudo gap model has been realized in \cite{Benini:2010qc,Chen:2011ny}. However, the order parameter field that couples to fermions to generate the gap in the fermionic spectral function is also the one characterizing the superconductor phase. As a result, when the order parameter vanishes, the gap disappears. Therefore, the pseudo gap phase appears at the temperature below the superconductor transition ($T* < T_c$). This problem can be solved in the expanse of introducing another field to generate a condensate that gives a gap to the fermion spectral function (analogous to \cite{Kiritsis:2015hoa}), while the condensate of another field is responsible for generating superconductivity. The explicit model was constructed as our model II. However, it would be more economic if we could use the field $\phi$ to do this job of generating a gap in the fermionic spectral function. Unfortunately, we have not succeeded in engineering a generic phase transition for $\phi$, otherwise using fields     $\phi$,  $\psi_1$ and $\psi_2$, we would be able to generate the anti-ferromagnetic, superconducting and pseudo gap phases using the set up of model I.

\subsection{Scaling symmetry}

Scaling symmetry is an important feature of phase diagram of a high temperature superconductor which suggests there is a quantum critical point hidden under the superconducting dome. The physics of the scaling symmetry can be understood from effective field theory. At the critical point, the theory is scale invariant such that the theory has no scale in the problem. Away from the critical point, the theory does not have exact scaling symmetry, but it is broken softly and the breaking can be computed in terms of the soft symmetry breaking parameter(s). Therefore, near the quantum critical point, the symmetry breaking at different points of phase space can be related by scaling. This is the scaling symmetry governed by the existence of quantum critical point.
A prime example of this type of scaling is the strange metal phase.  

\begin{figure}[H]
\begin{center}
\includegraphics[width=0.6 \textwidth]{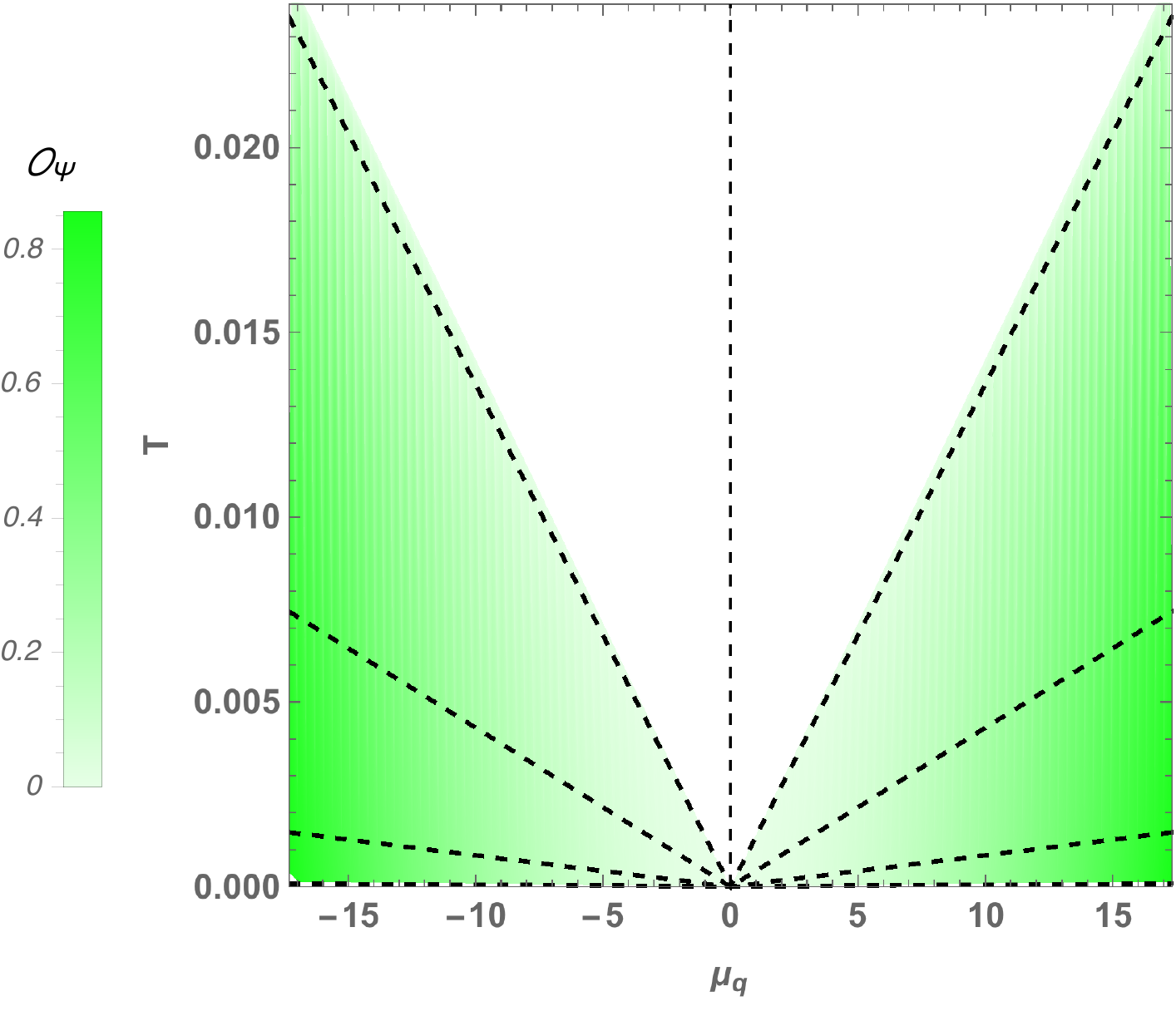}
\caption{The density plot of the order parameter $\cO_\Psi$ on the $\{\mu_q, T\}$ plane, which is equivalent to Figure \ref{figcon} of a single charged scalar field model in Appendix \ref{SectSSB}.
The transparent middle region has vanishing order parameter and is the normal phase, where the equation of state $\tilde\nN_{\nQ}\(\tilde\mu_\nQ\)$ in Equation \eqref{nqt} is independent of the temperature. There are two condensed phases; the right hand side one has a positive charge density $n_{\nQ}$, while the left hand side one has a negative $n_{\nQ}$. The dashed lines indicate the scaling trajectories. Note that this toy model has scaling symmetry even in the condensed phases. In this figure, we have set $g_F=1, L=1$ for convenience.
} \label{figchar}
\end{center}
\end{figure}

In AdS/CFT, a similar but even more power scaling symmetry is shown in the HHH model of holographic superconductors \cite{Hartnoll:2008vx}. As reviewed in Appendix \ref{SectSSB}, the order parameter versus temperature relation can be plotted in scaling invariant parameters 
 as in Figure \ref{figcon}. This plot is redrawn in Figure \ref{figchar} to demonstrate the scaling symmetry: once the physics on a constant $T$ slice is known, then the physics at all $T$ is known. The dashed lines on this plot shows physics on this line can be obtained by scaling. This is a nice demonstration that physics away from the quantum critical point at the origin is related to physics near the quantum critical point. In contrast to the scaling symmetry of a typical condensed matter system near a quantum critical point, not only the normal phase has the scaling symmetry, but also the condensed phase where symmetry is broken.     
 
The scaling symmetry in Figure \ref{figchar} seems to capture the essence of scaling symmetry in superconductors although it is known that the probe limit fails at zero temperature \cite{Gubser:2008wz,Gubser:2009cg} and some new ground state, such as the AdS soliton background \cite{Nishioka:2009zj,Horowitz:2010jq}, which is dual to an insulating phase, may appear and hence break the scaling symmetry\cite{Iqbal:2010eh,Horowitz:2009ij}. 

In our phase diagrams shown in Figures \ref{Fullphase2} and \ref{Fullphase1}, there is also scaling with respect to physics of different $\mu_q$ which has the same origin with the HHH model mentioned above which also has a powerful scaling symmetry for the condensates. However, the scaling symmetry in the $T$-$J_\phi$ plan is not manifest ($\mu_q$ is assumed constant on the phase diagram). It is possible by shrinking the superconducting dome to a point, certain scaling symmetry on the $T$-$J_\phi$ can emerge from the effective field theory argument. But apparently this cannot be seem easily through the bulk asymptotic equation of motion near the boundary. This will be further investigated in the future.

\section{Conclusion}
\label{SecCon}

We have studied how to engineer holographic models with features of a high temperature superconductor phase diagram. We introduced a field $\phi$ in the bulk which provides a tunable ``doping" parameter $J_\phi$ in the boundary theory. By designing how this field changes the effective masses of other order parameter fields, desired phase diagrams can be engineered. We have given examples of generating phase diagrams with phase boundaries similar to a superconducting dome and an anti-ferromagnetic phase by including two order parameter fields $\psi_1$ and $\psi_2$. We have also explored whether the pseudo gap phase can  be described without adding another order parameter field. This could have been achieved if the field $\phi$ can have a phase transition in the phase diagram. However, in our models, this was not realized. We have also examined how the scaling symmetry in the HHH model, which not only applies to the normal phase, but also the condensed phase arises. This symmetry naturally explains the scaling symmetry in the strange metal phase above the superconducting dome. However, this symmetry is not manifest in the $T$-$J_\phi$ phase diagram, instead, it related phase diagrams with different chemical potentials. This scaling symmetry could still appear if we study the light modes of the low energy effective field theory of the model. It is a topic for further investigation.

\section*{Acknowledgements}
 This work is supported by the MOST, NTU-CTS and the NTU-CASTS of Taiwan.

\appendix

\section{Free energy density}\label{probescalar}
 In this section, we calculate the
contribution to the free energy from the background gravity, the Maxwell field and the probe scalar fields. We will also follow this approach in Section \ref{SecPhase} for our model to understand the stability of all possible phases against each other.

In a grand canonical ensemble, the Gibbs free energy $\Omega$ is obtained from the partition function. In the AdS/CFT correspondence and the semi-classical limit, the partition function of the bulk theory is the path integral over the Euclidean metrics
\begin{align}
\Omega=T\log Z,\quad
Z=-\exp\(-S_{total}^{(E)}\),
\end{align}
where the Euclidean action is obtained via the Wick's rotation
\begin{align}
\tau=i t,\quad \tau\sim\tau+T^{-1}.
\end{align}

To fulfill our purpose, we mainly follow the reference \cite{Hartnoll:2009sz}. With a Dirichlet-like conformally flat boundary at infinity, the Euclidean action of the bulk Einstein-Maxwell theory is \cite{Henningson:1998gx,Balasubramanian:1999re,de Haro:2000xn}
\begin{align}\label{Eaction}
S^{(E)}_{\bg}&=- \int\dr r \int\dr \tau\dr^2 x \sqrt{g}\[\frac{1}{2\kappa^2}\(R+\frac{6}{L^2}\)-\frac{L^2}{2\kappa^2}\frac{1}{g^2_F}F^2\]+S_{\bt},
\end{align}
where $S_{\bt}$ contains the well-known Gibbons-Hawking term for the well-defined variational problem, and a constant boundary counter-term to cancel the divergence of the bulk action,
\begin{align}
S_{\bt}&=\int_{r\to\infty}\!\!\!\dr \tau\dr^2 x \sqrt{ h}\[\frac{1}{2\kappa^2}\(-2K+\frac{4}{L}\)+\frac{2L^2}{\kappa^2}\frac{\varepsilon}{g^2_F}n^aF_{ab}A^b\].
\end{align}
Here $h$ is the induced metric on boundary at $r \to \infty$, and $K=h_{\mu\nu}\nabla^{\mu} n^\nu$ is the trace of the extrinsic curvature of the boundary hypersurface, with $n^\mu$ an outward pointing unit normal vector.
$\varepsilon=0$ corresponds to the grand canonical ensemble with chemical potential $\mu_\nQ$ fixed, while $\varepsilon=1$ corresponds to the canonical ensemble with charge density $\nN_\nQ$ fixed.
We choose grand canonical ensemble ($\varepsilon=0$), and put the solutions \eqref{solution} into \eqref{Eaction} to obtain the on-shell Euclidean action  ${S^{(E)}_{\bg}}[g_{\mu\nu}, A_{\mu}]$. Then the free energy density turns out to be
\begin{align}
\frac{\Omega_0}{{\Vt}}=-\frac{T}{\Vt}\log Z_{\bg} =\frac{T}{{\Vt}} {S^{(E)}_{\bg}}[g, A_{\mu}]=-\frac{r_h^3}{2\kappa^2L^4}\(1+\frac{L^4}{g_F^2}\frac{\mu_\nQ^2}{r_h^2}\),
\end{align}
where $\Vt$ is the volume of the boundary system labelled by $(x, y)$.

When a system is in thermal equilibrium in a certain phase, the following thermodynamic relation is satisfied,
\begin{align}
\mathcal{E}+\mathcal{P}=Ts+\mu_\nQ\nN_\nQ.
\end{align}
And the energy density and pressure of a conformal matter are
\begin{align}
\mathcal{E}= \mathcal{T}^t_{~t},\quad \mathcal{P}=\mathcal{T}^x_{~x},\quad  \mathcal{E}=2\mathcal{P}.
\end{align}
Using the above thermodynamic relations, one obtains the free energy of the background gravity
\begin{align}
{\Omega_R}/{{\Vt}}\equiv \mathcal{E}- T s -\mu_\nQ\nN_\nQ = -\mathcal{P}.
\end{align}

Let us consider the simplest case for now.
The free energy contributed by a neutral scalar field $\Psi$ at the probe limit is
\begin{align}\label{Eaction2}
S^{(E)}_{\mM}&
= - \frac{{\Vt}}{T}\[ \int_{r_h}^{\infty}\dr r  \sqrt{g}  \mathcal{L}_{\Psi}+ \sqrt{\gamma}\mathcal{L}_{c.t.}\],\nn\\
g_M^2 \mathcal{L}_{\Psi}&=-\frac{1}{2}\(\partial\Psi \)^2-V(\Psi),\quad
V(\Psi)=\frac{1}{2}m_\Psi^2|\Psi|^2+\frac{1}{4}\mc_\Psi|\Psi|^4,
\end{align}
and $\mathcal{L}_{c.t.}$ is the boundary counter term to make the on-shell action
finite. The equation of motion for the neutral scalar field $\Psi$ turns out to be
\begin{align}\label{eompsi}
 r^{-2}   \partial_r\[r^4 f(r) \partial_r \Psi \]-m_{\Psi}^2 L^2\,\Psi
 -\mc_{\Psi} L^2 \Psi^3 = 0.
\end{align}
If we consider the contribution of the self interaction term $\lambda_\Psi |\Psi|^4$ in the Lagrangian density, the asymptotic behaviour of the scalar field receives non-trivial correction as follows,
\begin{align}
\Psi&\rightarrow \frac{J_{\Psi}}{r^{\Delta_{\Psi}^-} }+  \frac{O_{\Psi}}{r^{\Delta_{\Psi}^+} }
+\frac{\gamma }{r^{ 3 \Delta_{\Psi}^-}} +...,\quad
\gamma = \frac{\lambda_{\Psi} J_{\Psi}^3}{2\Delta_{\Psi}^-(4 \Delta_\Psi^- - 3)}, \label{phic}\\
 \Delta_{\Psi}^\pm&=\frac{3}{2}\pm \nu_{\Psi},\quad  \nu_{\Psi}=\sqrt{\frac{9}{4}+m^2_{\Psi} }.
\end{align}
In the case of $3\Delta_\Psi^- > \Delta_\Psi^+$, i.e. $m_\Psi^2<-27/16$, or $J_\Psi=0$,  the correction term in (\ref{phic}) is of higher order in $1/r$ and can be ignored, and the standard quantization scheme is not altered. However, for $3\Delta_\Psi^-\leqslant\Delta_\Psi^+$,
the counter term needs to be taken into account,
\begin{align}\label{Lagrangect}
g_M^2\mathcal{L}_{c.t.}&=\frac{\Delta_\Psi^-}{2 L} {\Psi}^2
+ \frac{\lambda_{\Psi} L}{4(4{\Delta_\Psi^-}-3)} {\Psi}^4 +...,
 \end{align}
to make the on-shell action finite at ultraviolet.

In more details,
after putting into the equation of motion \eqref{eompsi},
the on-shell formula of Lagrange density of the neutral scalar field $\Psi$ become
\begin{align}
\mathcal{L}_{\Psi}
&=-\frac{1}{2 g_M^2}\[\nabla_\mu\({\Psi}\nabla^\mu{\Psi}\)-\frac{1}{2}\mc_{\Psi}{\Psi}^4\]. 
\end{align}
Consider the background metric in \eqref{solution},
we obtain
\begin{align}\label{Lphi1}
- \int_{r_h}^{\infty}\dr r  \sqrt{g}\mathcal{L}_{\Psi}
&=\frac{1}{2g_M^2} \lim_{r\to\infty}  \[  \frac{r^4}{L^4}f(r)\({\Psi}\partial_r{\Psi}\)\]  - \frac{\mc_{\Psi}}{4g_M^2}\[\int_{r_h}^{\infty}\dr r \frac{r^2}{L^2}{\Psi}^4\].
\end{align}
Putting \eqref{phic} back into \eqref{Lphi1}, we have
\begin{align}
-\int_{r_h}^{\infty}\dr r  \sqrt{g}\mathcal{L}_{{\Psi}}
&=
\frac{1}{2g_M^2 L} \lim_{r\to\infty}  \frac{r^3}{L^3}\[-(\aa)^2{\Da}\(\frac{L}{r}\)^{2{\Dpa}}-3\aa\bb\(\frac{L}{r}\)^{3}
-3\aa\gamma{\Dpa}\(\frac{L}{r}\)^{4{\Dpa}}\]+\dots .
\end{align}
Following the procedure of holographic renormalization~\cite{Skenderis:2002wp},
we need to introduce the counter terms in \eqref{Lagrangect}. 
The leading order expansions in the action are
\begin{align}
\frac{{\Da}}{2 L}\int  \sqrt{h}{\Psi}^2&=\frac{1}{2 L}\lim_{r\to\infty}
\frac{r^3}{L^3}\[(\aa)^2{\Da}\(\frac{L}{r}\)^{2{\Da}}+2{\Da}\aa\bb\(\frac{L}{r}\)^{3}+2\aa\gamma{\Da}\(\frac{L}{r}\)^{4{\Da}}\]+\dots\nn\\
 \frac{\lambda_{\Psi}L}{4(4{\Da}-3)}
\int  \sqrt{h}{\Psi}^4&=\frac{1}{2 L}\lim_{r\to\infty}
\frac{r^3}{L^3}\[+\aa\gamma{\Dpa}\(\frac{L}{r}\)^{4{\Da}}\]+\dots
\end{align}
Put all of these terms into the total action \eqref{Eaction2},
we reach the finite formula of the on shell Euclidean action.
As a result, the free energy density becomes
\begin{align}\label{Fphi}
\frac{\Delta\Omega_\Psi}{{\Vt}} &= \frac{1}{g_M^2}\[-\nu_{\Psi} J_{\Psi} O_{\Psi}
-\frac{\lambda_{\Psi}}{4}\int_{r_h}^{\infty}\dr r \frac{r^2}{L^2}\({\Psi}^4\)
-\lim_{r\to\infty} \frac{ \lambda_{\Psi} J_{\Psi}^4 \, L}{4 (4 \Delta_\Psi^- -3)}\(\frac{L}{r}\)^{4 \Delta_\Psi^- - 3}\].
\end{align}
To keep our computation simple, we only consider the scalar field mass $m_\Psi^2<-27/16$ or set $\lambda_\Psi=0$.
Therefore the last term in \eqref{Fphi} drops out.

\section{Scaling symmetry in the HHH model \cite{Hartnoll:2008vx}}
\label{SectSSB}

\begin{figure}[H]
\begin{center}
\includegraphics[width=0.6\textwidth]{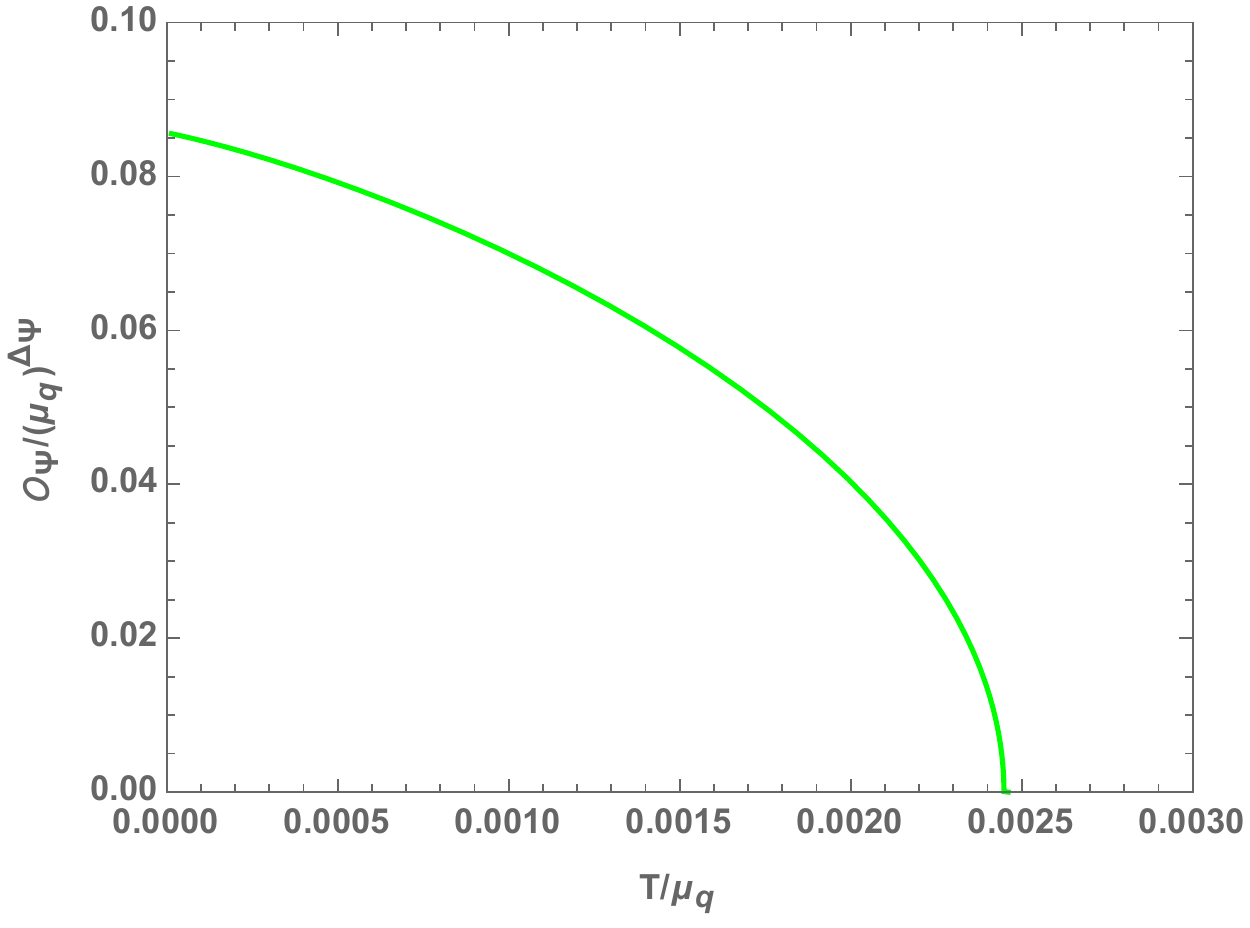}
\caption{The plot of expectation value $\cO_{\Psi}$ versus the temperature $T$, at fixed chemical potential $\mu_q$. Both of $\cO_{\Psi}$ and $T$ are rescaled by appropriate power of chemical potential $\mu_q$. The following parameters ate used: $J_\Psi=0, m_\Psi^2L^2=-2.1, \mc_\Psi=1$, and $q_\Psi=0.5$.}
\label{figcon}
\end{center}
\end{figure}

Here we review a powerful scaling symmetry shown in Figure \ref{figcon} based on the HHH model \cite{Hartnoll:2008vx}. In this model, not only the unbroken phase in Figure \ref{figcon} but also the symmetry breaking phase enjoy a scaling symmetry. This is more powerful than a typical field theory system with a quantum critical point where only the unbroken phase enjoys the scaling symmetry. 

Consider a massive $U(1)$ charged scalar field $\Psi$ coupled to a Maxwell field $A$ in the gravitational background of Equation(\ref{action}). The dimensionless coupling is $q_{\Psi}$. To initiate the spontaneous symmetry breaking, it generally requires a non-trivial potential for $\Psi$. The simplest form is the Higgs-like potential $V(\Psi)$,
\begin{align}\label{singles}
g_M^2 \mathcal{L}_{M} = g_M^2 \mathcal{L}_{\Psi}&=-\frac{1}{2}|\partial\Psi- i q_\Psi A\Psi |^2-V(\Psi),\nn\\
V(\Psi)&=\frac{1}{2}m_\Psi^2|\Psi|^2+\frac{1}{4}\mc_\Psi|\Psi|^4.
\end{align}
(In the model of Ref. \cite{Hartnoll:2008vx}, $\mc_\Psi=0$.)
For simplicity, we study the phase transition of $\Psi$ in the probe limit, namely ${2\kappa^2}/{g_M^2}\rightarrow0$, such that the energy density of the fluctuations in $\Psi$ is very small compared to that of the background. The symmetry broken phase boundary is obtained as the onset of the condensation of $\Psi$ in the probe limit of the charged black brane background. The equation of motion is 
\begin{align}\label{EOMcharge}
\frac{1}{r^2 L^2}  \partial_r\[r^4 f(r) \partial_r{\Psi}\]-\(m_\Psi^2-\frac{L^2q_\Psi^2{A_t(r)}^2}{r^2f(r)} \)\Psi-\mc_\Psi|\Psi|^2\Psi=0.
\end{align}
Note that near the boundary, the last two terms are negligible compared with the first two terms. This implies a bigger symmetry near the boundary than near the black hole horizon. 

Near the boundary, $\Psi$ has the asymptotic behaviour:
\begin{align} \label{assympsi}
\Psi\rightarrow \frac{J_\Psi}{r^{\Delta_{\Psi}^-}}+ \frac{\cO_\Psi}{r^{\Delta_{\Psi}^+}}+...,\qquad  \Delta_{\Psi}^\pm=\frac{3}{2}\pm\sqrt{\frac{9}{4}+m^2_{\Psi} L^2},
\end{align}
where the standard quantization identifies $J_\Psi$ as the source and $\cO_{\Psi}$ as the vacuum expectation value (VEV) in the dual boundary theory. Under the $r \to \lambda r $ scaling of Equation(\ref{scalingx}), the asymptotic equation of motion remains invariant under the tranformation:
\begin{align} \label{xxx}
J_\Psi\rightarrow \lambda^{\Delta_{\Psi}^-} J_\Psi, \qquad \cO_\Psi \rightarrow \lambda^{\Delta_{\Psi}^+} \cO_\Psi. 
\end{align}
This implies once the  $\cO_{\Psi}$ versus $T$ relation is known at certain chemical potential $\mu_q$, its relation will be known to all $\mu_q$ as well.

Therefore, we can plot Figure \ref{figcon} in scaling invariant coordinates. Or we can replot it as Figure \ref{figchar} to show the scaling symmetry more explicitly. This scaling symmetry is so powerful that it involves not only the normal phase but also the symmetry breaking phase, which is not usually seem in condensed matter systems.

\end{document}